\documentclass{mem}
\usepackage{natbib}\usepackage{txfonts}\usepackage{balance}
\usepackage{graphicx}
\begin{document}
\def\teff{$T\rm_{eff }$}
\def\kms{$\mathrm {km s}^{-1}$}

\newcommand{\rc}[1]{{\textcolor{blue}{#1}}}

\title{3D Hydrodynamical Simulations of a Brown Dwarf Accretion by a Main-Sequence Star and  its Impact on the Surface Li Abundance
}

   \subtitle{}

\author{
C. \,Abia\inst{1},
R. M. \,Cabez\'on\inst{2},
\and I. \,Dom\'\i nguez\inst{1}
}

\institute{ Dpto. F\'\i sica Te\'orica y del Cosmos,
Universidad de Granada, E-18071 Granada, Spain
\email{cabia@ugr.es}
\and
Scientific Computing Core (sciCORE), Universit\"at Basel, Klingelbergstrasse 61, 4056 Basel, Switzerland
}

\authorrunning{Abia, Cabez\'on, Dom\'\i nguez}

\titlerunning{3D hydrodynamical simulations}

\abstract{Li-depleted/enhanced stars in the main-sequence (MS) and/or the RGB, pose a puzzling mystery. Presently, there is still no clear answer to the mechanism(s) that enables such Li depletion/enhancement. One possible explanation comes from the - still controversial – observational evidence of Li underabundances in MS stars hosting planets, and of a positive correlation between the Li abundance and rotational velocity in some RGB stars, which suggests a stellar collision with a planet-like object as a possible solution. In this study we explore this scenario, performing for first time 3D-hydrodynamical simulations of a 0.019~M$_\odot$ brown dwarf collision with a MS star under different initial conditions. This enables us to gather information about the impact on the physical structure and final Li content in the hosting star. 

\keywords{Methods: numerical --
Planet-star interactions -- Stars: low mass}
}
\maketitle{}

\section{Introduction}

The accretion of planet-like material onto stars have been considered as a possible explanation of the Li abundance anomalies found in main-sequence (MS) and red giant stars. Jupiter-like planets and
brown dwarfs (BD) are expected to preserve their original Li content. However,  Li is a very fragile element that is rapidly destroyed in stellar interiors by proton capture reactions at T$\ga 2.5 \times 10^6$ K. Since these temperatures are easily reached in the interior of low mass stars in the MS, Li is one of the best tracers of internal mixing mechanism and of possible associated angular momentum transport. For cool MS stars, the best constraints are the Solar Li abundance, which is a factor $\sim 150$ lower than the meteoritic value \citep{lod03}, and the Li abundance found in solar twins \citep[e.g.][]{mai19}. A large range of Li abundances is observed in solar-type stars of very similar age, mass and metallicity as the Sun, but such a range is theoretically difficult to understand. Rotational-induced processes coupled to the effects of internal gravity waves, atomic diffusion, mass loss, and magnetic instabilities are the suggested  mechanisms able to provide a consistent explanation for the observed behaviour of the surface Li abundance in these MS stars, although difficulties still remain \citep[see, e.g.][]{mic86,cha05,den10,egge12,cha19}. After the discovery of extra-solar planets \citep{may95}, the interaction planet-star was added to the above list of mechanisms to explain the Li abundance anomalies observed in some MS and red giant stars. 
\citet{isr09} suggested that the presence of planets may increase the amount of mixing and deepen the convective zone of MS stars in such extend that Li can be more extensively burned. Direct engulfment of a planet-like object may increase this effect. Current Li abundance surveys in MS stars are not conclusive on the fact whether stars hosting planets show different Li abundances than those which (apparently) do not host planets. The reason of this is, in part, the difficulty in finding a large enough homogeneous sample of stars having equal age, mass, metallicity, rotational velocity, etc., since Li depletion in MS stars is very sensitive to all these parameters.   

To enlighten this problem, here we show preliminary 3D hydrodynamical simulations of the collision between a $\sim 1$ M$_\odot$ MS star and a $\sim 0.019$ M$_\odot$ ($\sim 20$ M$_J$) BD. Both objects are assumed to have solar metallicity. Our aim is to gather information about the impact on the physical structure and final Li content in the hosting star.

\begin{figure*}[t!]
\resizebox{\hsize}{!}{\includegraphics[clip=true]{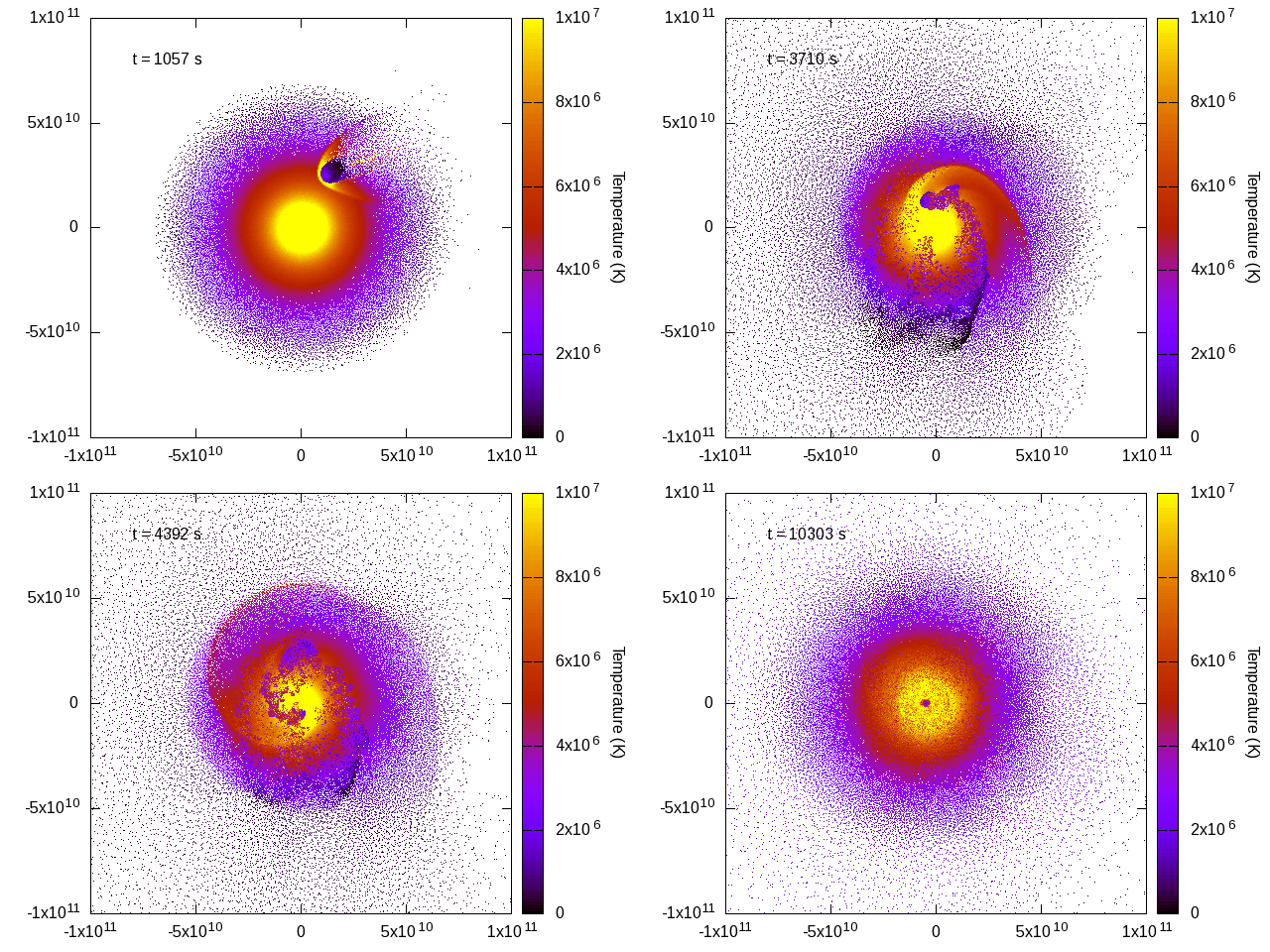}}
\caption{Thin slice at the orbital plane of the 3D particle distribution depicting the evolution of case C at different times (labelled). The temperature in K is color-coded. Each snapshot shows a box of $2\times 10^{11}$ cm side.}
\label{fig1}
\end{figure*}

\section{3D SPH simulations}
Our 3D simulations use the Smooth Particle Hydrodynamics (SPH) method. An up-dated description can be see in \citet{rosswog2015}. In particular, we use SPHYNX\footnote{https://astro.physik.unibas.ch/sphnyx} code, which is a state-of-the-art SPH code. Technical details can be seen in \citet{cab17}. The calculations were performed in the sciCORE (Basel) and Alhambra (Granada) supercomputer facilities using about 150,000~CPUh per simulation. The total simulated physical time was $\sim 16,000$~s. 

The initial structure of the MS star was chosen to be similar to the current Sun following \citet{bac05}, while that of the BD was built using a polytrope of index 1.5 according to \citet{bur01}. The number of SPH particles were $\sim 5\times 10^6$ and $10^5$ for the MS and BD, respectively, with $\sim 2\times 10^{-7}$ M$_\odot$ per particle in both cases. Four collision scenarios
were simulated: A) direct collision (impact parameter $b=0$), B) a collision with an impact parameter $b\sim 0.5$~R$_\odot$, C) the same as case B but including a initial rotational velocity of 2 and 25~km~s$^{-1}$ for the MS star and BD, respectively and, D) a smoother spiral-in merging. At the moment of this presentation, the numerical simulations of the three first cases were finished while case D was in progress. Case A is a limiting case and we used it for testing and calibration purposes. Here we describe briefly the main results of case C; a full discussion will be presented elsewhere. The case of the interaction between a red giant star and a planet-like object will also be addressed in a future study.

\section{Results}
In Fig.~1, we show four snapshots of a thin cut along the equatorial plane, which is also the plane of the collision. The temperature is color-coded. Initially, (t~$\lesssim 2,000$ s) the BD falls in at a supersonic velocity ($\mathcal{M}\sim 14$ at first contact), producing a well-defined bow shock wave at its front (top left panel). For the next few thousands of seconds, the BD follows a spiral-in trajectory still supersonically ($\mathcal{M}\sim 4-2$), but decreasing its speed due to the increasing drag when traversing denser regions of the MS. During this phase the BD compresses significantly, losing material to the surrounding MS layers via ablation. Only for t~$\gtrsim 4,000$ s, after one spiral-in orbit, the BD's velocity is lower than the local sound speed. This triggers a great amount of hydrodynamical instabilities that enhance the mixing of the BD material with its surroundings. Nevertheless, this stage happens when the BD is already deep inside the MS, avoiding a significant dilution of the BD material in the outer convective layers of the MS star. In fact, most of the BD mass will reach the radiative core of the host star (bottom left and right panels). This result is at odds with that assumed in previous 1D simulations \citep[e.g.][]{san98,mon01}. On the other hand, at the end of the simulation it is found a significant drop in the central temperature (by a factor $3-5$) with respect to the initial value (bottom right panel) and an increase in the central density by a factor $3-9$, depending on the radial direction. This shows that the engulfment initially produces a rather {\it asymmetric object}. The collision also induces a {\it fast differential rotation} in the final object: the core rotates with a period $\sim 2$~h, while the average rotational period is $\sim 18$~h (i.e. a factor $\sim 30$ shorter than initial period). Nevertheless, the corresponding angular velocities are still below the critical value. Differential rotation and a fast rotating core is also produced in cases A and B. This is encouraging since there is observational evidence that planet engulfment may induce a significant increase of the rotational velocity of the host star \citep[at least in red giants, e.g.][]{car13}, although not as large as we find here. Note, however, that cases discussed here are not probably realistic engulfment scenarios; case D (in progress) simulates a smoother merging between both stars, being a more realistic process.

From Fig.~1 is also evident that the final outcome results in a considerable expansion of the host star. Since this expansion is not spherically symmetric this makes difficult to estimate the final radii. 
Note that the outer parts of the star at the end of the simulation are not well defined because of the scarce number of mass particles outside a given radius. Having said that, a crude estimate shows that the radii may expand by a factor of $\sim 3$. However, note also that at the time at which the simulation was stopped the system was not completely stabilised, therefore, we can not discard a final shrink of the whole star when the outer layers, still gravitationally bound, fall back onto the star. 

Another interesting outcome is the mass loss during the collision. A straightforward estimate can be done considering the number of particles moving with a radial velocity higher than the escape velocity at their radius, at the end of the simulation. Doing this we obtain in case C that a mass $\sim 7 \times 10^{-4}$~M$_\odot$ is lost, a rather considerable amount of mass. Note for instance, that this mass would represent~$\sim 60\%$ of the current planetary mass (planets, comets, asteroids etc) in the Solar system. This result is also encouraging since a significant fraction of red giants with infrared excess, usually associated to the existence of circumstellar mass, show Li enhancements, which could be produced by a previous planet-like engulfment \citep[e.g.][]{zho19}.  

Additionally, we can estimate the Li variation produced in the surface of the MS star by the BD engulfment. To do this we consider only the Li mass of the BD deposited above the convective envelope of the host star. We assume that any Li mass of the BD not deposited in the convective region will be finally destroyed because the temperature there will exceed $\sim 2.5\times 10^6$~K, or in other case, because actually can not be observed not being in the convective region. In case C, the final mass distribution indicates that only $\sim 6\times 10^{-7}$ M$_\odot$ of the BD mass would be diluted in the convective zone. Assuming an initial surface Li abundance in the MS star similar to that in the present Sun ($\sim 1.05$)\footnote{Li abundances are given in logarithmic scale where log N(H)$\equiv 12$.}, and a meteoritic Li abundance  ($\sim 3.3$) in the BD, we obtain that the surface Li abundance would increase by only 0.03 dex (i.e. $\sim 1.08$), which can be hardly detected considering the current observational uncertainties in the derivation of Li abundances in MS stars. 

Finally, it would be interesting to follow the subsequent evolution of the final structure. The new star is denser and cooler at the centre, it is more expanded and a fast rotator. Obviously this should have important consequences on the evolution of the merged object.  To do that, first the 3D structure should be mapped into a 1D structure suitable for a 1D hydrostatic evolutionary code. However, in this way the resulting asymmetry in the new stellar object would be lost, and it is expected that, to conduct the convergence of the 1D model, some properties have to be modified.

\begin{acknowledgements}
This study has been partially supported by the Spanish Grants AYA2015-63588-P and PGC2018-095317 within the European Funds for Regional Development (FEDER) and the Swiss Platform for Advanced Scientific
Computing (PASC) via the “SPH-EXA: Optimizing Smoothed Particle Hydrodynamics for Exascale Computing” project. We also acknowledge the support of the scientific computing center at University of Basel (sciCORE) and the Centro de Servicios Inform\'aticos y Redes de Comunicaciones (CSiRC) at the University of Granada, where these calculations where performed.
\end{acknowledgements}

\bibliographystyle{aa}

\end{document}